\documentclass[aps,prl,reprint,showpacs,amsmath,superscriptaddress]{revtex4-1}

\usepackage{amssymb}
\usepackage{mathrsfs}
\usepackage{graphicx}
\usepackage[normalem]{ulem}
\usepackage{color}
\usepackage{bm}

\begin{document}

\title{Floquet higher order topological insulator in a periodically driven bipartite lattice}
\author{Weiwei Zhu}
\affiliation{Department of Physics, National University of Singapore, Singapore 117542, Singapore}

\author{Y.~D.~Chong}
\email{yidong@ntu.edu.sg}
\affiliation{Division of Physics and Applied Physics, School of Physical and Mathematical Sciences, Nanyang Technological University, Singapore 637371, Singapore}
\affiliation{Centre for Disruptive Photonic Technologies, Nanyang Technological University, Singapore 637371, Singapore}
\author{Jiangbin Gong}
\email{phygj@nus.edu.sg}
\affiliation{Department of Physics, National University of Singapore, Singapore 117542, Singapore}

\begin{abstract}
  Floquet higher order topological insulators (FHOTIs) are a novel topological phase that can occur in periodically driven lattices.  An appropriate experimental platform to realize FHOTIs has not yet been identified.  We introduce a periodically-driven bipartite (two-band) system that hosts FHOTI phases, and predict that this lattice can be realized in experimentally-realistic optical waveguide arrays, similar to those previously used to study anomalous Floquet insulators.  The model exhibits interesting phase transitions from first-order to second-order topological matter by tuning a coupling strength parameter, without breaking lattice symmetry.  In the FHOTI phase, the lattice hosts corner modes at eigenphase $0$ or $\pi$, which are robust against disorder in the individual couplings.
\end{abstract}

\maketitle
{\it Introduction.}---Nonequilibrium topological matter generated by periodic driving has been a frontier research topic during the past decade \cite{Rudner2020Rev}.  Periodically-driven or ``Floquet'' systems can realize exotic topological phases not found in static systems, such as Floquet $\pi$ modes \cite{Jiang2011pimode,Tong2013pimode}, anomalous Floquet insulators (AFIs) \cite{Rudner2013AFI}, and space-time symmetry protected topological insulators \cite{Morimoto2017spacetimesymmetry}.  This has motivated theoretical and experimental studies of light-irradiated materials \cite{Kitagawa2010Grapene,Kitagawa2011Grapene,Mclver2020Grapene,OKa2009Grapene,OKa2009Grapene,NC2017} as well as driven cold-atom systems \cite{Wintersperger2013coldatom}.  Floquet topological phases can also be realized in time-independent platforms such as coupled optical resonators or oriented scattering networks \cite{Liang2013CRR, Pasek2014CRR, Hu2015CRR, Gao2016CRR, Delplace2017CRR, Afzal2020CRR} and optical waveguide arrays \cite{Leykam2016waveguide, Maczewsky2017waveguide, Maczewsky2020waveguide, Mukherjee2017waveguide, Mukherjee2018waveguide, Mukherjee2020waveguide, Rechtsman2013waveguide}, in which the process of scattering or wave propagation simulates the time evolution of a wavefunction.

Higher order topological insulators (HOTIs) are an intriguing group of topological phases \cite{Benalcazar2017HOTI,Schindler2018HOTI} that feature gapped first-order boundary and gapless higher-order boundary.  Soon after the discovery of HOTIs, several groups have investigated the possibility of realizing Floquet HOTIs (FHOTIs) \cite{Bomantara2019FHOTI, Vega2019FHOTI1, Nag2019FHOTI1, Seshadri2019FHOTI1, Peng2019FHOTI2, Peng2020FHOTI2, Chaudhary2020FHOTI2, Ghosh2020FHOTI1, Hu2020FHOTI1, Huang2020FHOTI1, Bomantara2020FHOTI, Gong2020, Saha2020, Zhang2020FHOTI}, such as the use of driving schemes whose instantaneous Hamiltonians possess the symmetries of static HOTIs \cite{Ghosh2020FHOTI1, Hu2020FHOTI1, Huang2020FHOTI1, Nag2019FHOTI1, Seshadri2019FHOTI1, Vega2019FHOTI1}, as well as the use of peculiar space-time symmetries that are unique to periodically driven systems \cite{Chaudhary2020FHOTI2,Peng2019FHOTI2,Peng2020FHOTI2}.  To date, it remains unclear what is the ideal experimental platform for realizing a FHOTI and studying its properties.

Here, we present a periodically-driven bipartite square lattice model hosting FHOTI phases that should be easily experimentally accessible.  Unlike previous FHOTI proposals that have required either negative hopping/coupling \cite{Hu2020FHOTI1,Huang2020FHOTI1},  spin-orbital or superconducting interactions \cite{Vega2019FHOTI1, Seshadri2019FHOTI1, Nag2019FHOTI1, Peng2019FHOTI2, Ghosh2020FHOTI1, Peng2020FHOTI2, Chaudhary2020FHOTI2, Gong2020, Saha2020, Bomantara2020FHOTI, Zhang2020FHOTI}, our model involves a simple two-band single-particle Hamiltonian with only periodic time modulation in the on-site potential differences and nearest-neighbor couplings, and with all couplings strictly non-negative throughout the driving protocol.  These features allow the model to be implemented in experimental platforms like optical waveguide arrays \cite{Leykam2016waveguide, Maczewsky2017waveguide, Maczewsky2020waveguide, Mukherjee2017waveguide, Mukherjee2018waveguide, Mukherjee2020waveguide, Rechtsman2013waveguide} and coupled optical resonator lattices \cite{Liang2013CRR, Pasek2014CRR, Hu2015CRR, Gao2016CRR, Delplace2017CRR, Afzal2020CRR}---i.e., minor variations of the experimental setups previously used to realize Floquet topological insulators \cite{Leykam2016waveguide, Maczewsky2017waveguide, Mukherjee2017waveguide, Mukherjee2018waveguide, Maczewsky2020waveguide}. 
{Furthermore, because the system has only two bands,  it becomes possible to excite the edge mode or corner mode by a single-site source, thus simplifying experimental procedures in probing and distinguishing between bulk, edge, and corner modes.}  The model proposed here thus offers a highly promising route to the experimental realization of a FHOTI.

\begin{figure}
  \includegraphics[width=1\linewidth]{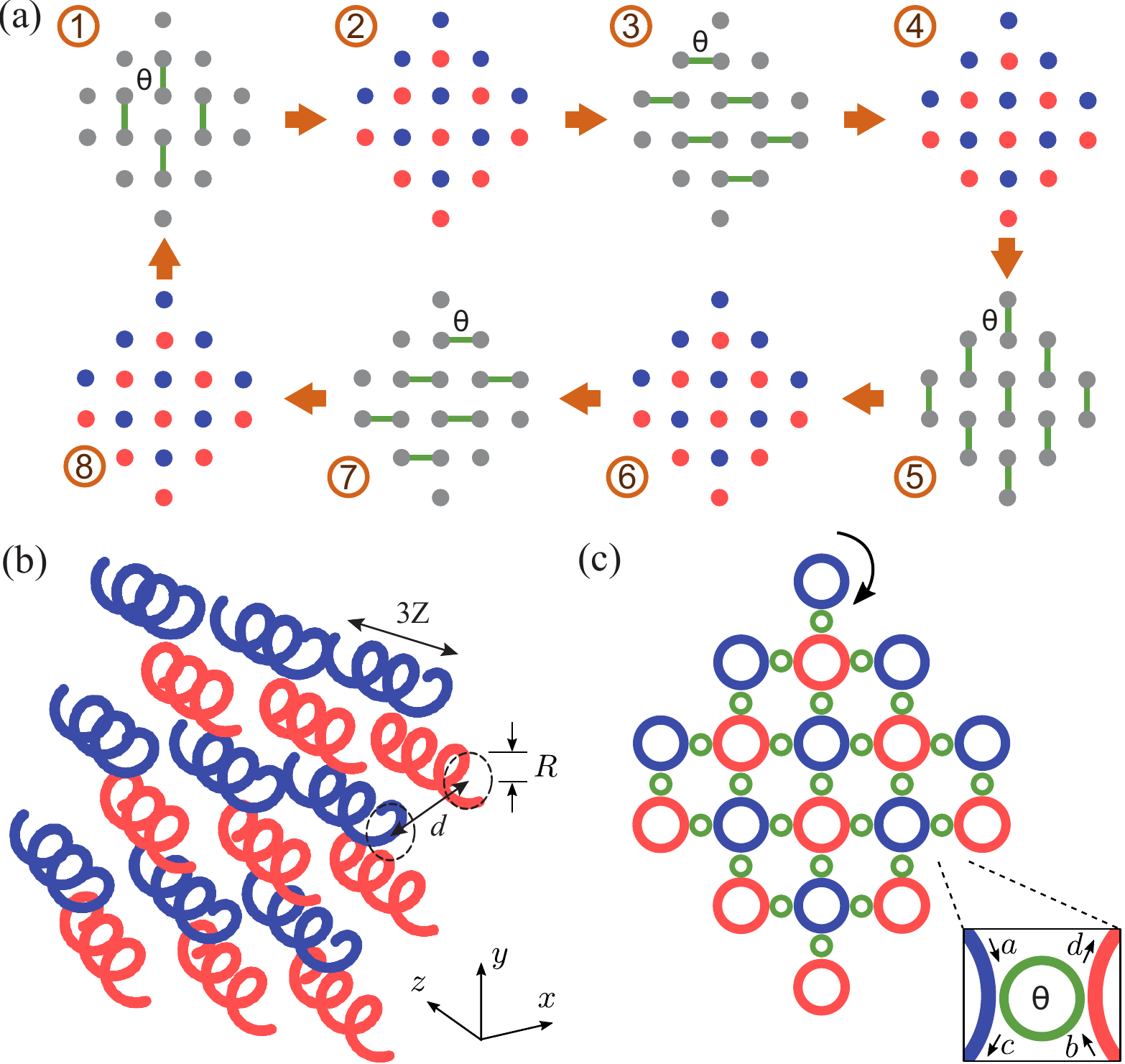}
  \caption{Periodic modulation protocol for realizing a Floquet high-order topological insulator (FHOTI).  (a) Driving protocol for sites in a square lattice.  During even time steps, the sites are decoupled and the two sublattices (red and blue circles) experience an energy bias $\pm \Delta$.  During each odd time step, one of the four possible nearest neighbor couplings is activated with hopping strength $\theta$. (b) Waveguide array equivalent to (a) for a given choice of axial ($z$) propagation direction.  The waveguides are helical, with neighboring waveguides staggered by a half period along $z$.  (c) Lattice of coupled ring resonators, which is equivalent to (a) for a choice of light propagation direction within the rings.  Nearest neighbor site rings (red and blue circles) are coupled by off-resonant auxiliary coupling rings (green circles).  }
  \label{diagram}
\end{figure}

{\it Model.}---Consider a tight-binding model of a bipartite square lattice containing periodically-modulated stepwise nearest neighbor couplings.  As shown in Fig.~\ref{diagram}(a), the modulation consists of 8 steps of equal duration $T/8$, where $T$ is the modulation period.  At odd steps, one of the four sets of nearest neighbor couplings are activated (so instantaneous system is dimerized).  At even steps, the two sublattices $A$ and $B$ (red and blue) experience a potential difference.  The time-dependent Bloch Hamiltonian thus becomes
\begin{equation}\label{eq1}
  H(\mathbf{k},t) = \!\sum_{m=1,3,5,7} \! \theta_{m}(t)
  \left(e^{i \mathbf{b}_{m}\cdot \mathbf{k}}\sigma^{+} + \textrm{h.c.}\right)
  + \Delta(t)\sigma_{z},
\end{equation}
where $\theta_{m}(t)$ is set to be $\theta$ during the $m$-th (modulo 8) step, and zero at other time steps; $\Delta(t)$  equals $\Delta$ at even steps, and zero at odd steps.  $\sigma^{\pm}=(\sigma_{x}\pm i\sigma_{y})/2$, where $\sigma_{x,y,z}$ are Pauli matrices; and the vectors $\mathbf{b}_{2n+1}$ are given by $-\mathbf{b}_{1}=\mathbf{b}_{5}=(0,a)$ and $-\mathbf{b}_{3}=\mathbf{b}_{7}=(a,0)$, where $a$ is the lattice constant.  For the purposes of theoretical modelling, we always assume $T=8$ and $a=1$.

Using the instantaneous Hamiltonian, we define the Floquet operator (i.e., the time evolution operator over one period), $U_F \equiv \mathcal{T}\mathrm{exp}\big[-i\int_{t_0}^{t_0+T} H(\tau)\,d\tau\big]$, where $\mathcal{T}$ is the time-ordering operator and $t_0$ is a reference time.  We identify solutions $|\psi(t)\rangle=e^{-i\varepsilon t}|\phi(t)\rangle$, with $|\phi(t)\rangle=|\phi(t+T)\rangle$ and
\begin{equation}\label{eq2}
  U_F\, |\phi(t_0)\rangle = e^{-i\varepsilon T} |\phi(t_0)\rangle.
\end{equation}
The quasienergy $\varepsilon$ is an angular variable with period $2\pi/T$. Floquet states with $\epsilon =\pm \pi/T$ are called $\pi$ modes in the literature because their eigenphase is $\pi$.

The model proposed above can be realized using an optical waveguide array, as depicted in Fig.~\ref{diagram}(b), or a coupled optical ring resonator lattice, as shown in Fig.~\ref{diagram}(c).  For the waveguide array, the propagation of the paraxial light field envelope maps onto a time-dependent Schr\"odinger equation; the helical waveguides at the $A$ and $B$ sites are staggered along the axial direction, which reproduces the stepwise activation of the inter-site couplings \cite{Leykam2016waveguide, Maczewsky2017waveguide, Maczewsky2020waveguide, Mukherjee2017waveguide, Mukherjee2018waveguide, Mukherjee2020waveguide, Rechtsman2013waveguide} and hence yields the time-dependent Hamiltonian~\eqref{eq1}.  In the coupled-ring platform, fixed-frequency light waves propagate within a lattice of ``site rings'' with a given circulation direction (i.e., either clockwise or anticlockwise within each ring, with no back-propagation); by accounting for the unitary scattering relations between the coupled rings, one can derive a spectral-band problem equivalent to the Floquet band eigenproblem of Eq.~\eqref{eq2} \cite{Pasek2014CRR, Delplace2017CRR}.

\begin{figure}
\includegraphics[width=1\linewidth]{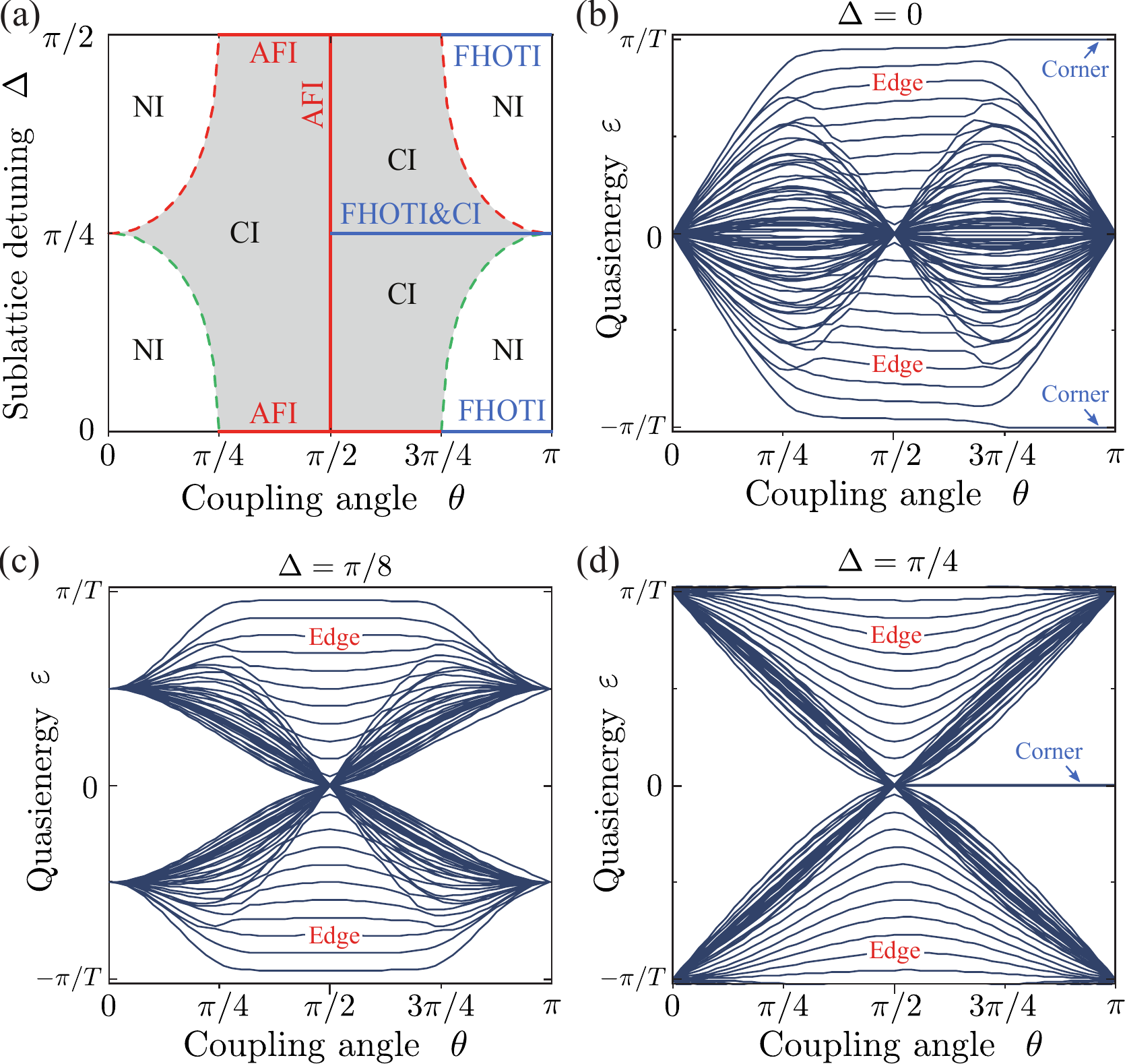}
\caption{(a) Phase diagram of the lattice, plotted versus the coupling angle $\theta$ and sublattice detuning $\Delta$. Green (red) dashed lines are the phase transition point at $\Gamma$ (X) point. (b)--(d) Spectrum as a function of coupling strength $\theta$ for a finite structure with 8 unit cells along each edge direction. (b) for $\Delta=0$, (c) for $\Delta=\frac{\pi}{8}$ and (d) for $\Delta=\frac{\pi}{4}$.}
\label{square}
\end{figure}

{\it Topological phases.}---Fig.~\ref{square}(a) shows the obtained topological phase diagram, plotted against the potential bias parameter $\Delta$ and the coupling strength parameter $\theta$.  A quick inspection of Fig.~\ref{square}(a) shows that several distinct topological phases are induced by periodic time modulation: normal insulator (NI), Floquet Chern insulator (CI), AFI \cite{Rudner2013AFI}, and FHOTI.  At the phase boundaries, the band-crossing points occur at either the $\Gamma$ or $X$ points in the Brillouin zone, respectively indicated by green and red dashes in Fig.~\ref{square}(a).  The band-crossings occur at these points because the Hamiltonian~\eqref{eq1} is invariant under rotation by $90$ degrees ($C_{4}$) and a quarter-period translation in time: $H(\mathbf{k},t)=H(C_{4}^{-1}\mathbf{k},t+T/4)$, so that $U_F(\mathbf{k}) = U_F(C_{4}^{-1}\mathbf{k})$.  The AFI phases (in which the Floquet band-structure has a single nontrivial bandgap) exist only along the red lines in Fig.~\ref{square}(a). By constract, the FHOTIs exist only along the shown blue lines, requiring $\Delta = 0$ (same as $\Delta = \pi/2$) or $\Delta = \pi/4$.  The found FHOTI phases are protected by a combination of particle-hole symmetry and inversion symmetry, as explained below.

The band diagram for $\Delta = 0$ is shown in Fig.~\ref{square}(b).  In this case, there is a single bulk band, and the range of $\varepsilon$ outside the band consitutes a single ``band gap'' (recall that $\varepsilon$ is a periodic variable).  For $0 \le \theta < \pi/4$, the gap is topologically trivial; this is the NI phase.  For $\pi/4 < \theta < 3\pi/4$, the gap is topologically nontrivial and spanned by edge modes; this is the AFI phase \cite{Rudner2013AFI, Liang2013CRR, Pasek2014CRR, Delplace2017CRR}.  For $3\pi/4 < \theta < \pi$, the system is in the FHOTI phase with corner modes at $\varepsilon = \pi/T~(-\pi/T)$.

To intuitively understand the emergence of Floquet corner modes, it is useful to tentatively use the coupled ring lattice representation \cite{Pasek2014CRR, Delplace2017CRR}.  The stepwise inter-site couplings in the Floquet system of Fig.~\ref{diagram}(a) are equivalent to a set of unitary scattering relations.  As shown in Fig.~\ref{diagram}(c), these relations have the form $[c,~d]^T = \exp(-i\theta\sigma_x)\,[a,~b]^T$, where $\{a,b,c,d\}$ are complex wave amplitudes in the incoming and outgoing arms of adjacent rings.  For $\theta = \pi$, the scattering relation reduces to $-a\rightarrow c$ and $-b\rightarrow d$; the rings effectively become decoupled, with the inter-ring connection only contributing a $\pi$ phase shift with each ring.  Under these circumstances, one round of propagation within a bulk ring (which has four neighbors) or edge ring (which has two neighbors) acquires an even number of $\pi$ phase shifts; however, one round around a corner ring (which has one neighbor) brings a $\pi$ phase shift.  This implies that there exist Floquet eigenstates with $\varepsilon T = 0$ localized to each bulk and edge site, but a separate set of Floquet eigenstates with $\varepsilon T = \pi$ localized to each corner site.  For $\theta < \pi$, the corner modes persist so long as the gap remains open.

Fig.~\ref{square}(c) shows the band diagram for $\Delta=\pi/8$.  Here, the system supports NI, AFI and CI phases.  The AFI phase exists only at $\theta = \pi/2$, when both bulk bands collapse to $\varepsilon = 0$.
Fig.~\ref{square}(d) shows the band diagram for $\Delta=\pi/4$.  This is the other scenario in which corner modes appear.  For $0 < \theta < \pi/2$, the system is in an AFI phase with two bulk bands, a trivial gap, and a nontrivial gap spanned by edge modes.  For $\pi/2 < \theta \le \pi $, the system features the coexistence of FHOTI and CI phases.  There are two bulk bands, a gap spanned by edge modes, and a gap containing a corner mode pinned at $\varepsilon = 0$.

{\it Symmetry analysis and robust of FHOTIs.}---The corner modes are seen to only appear at $\Delta=n\pi/4$ where $n\in \mathbb{Z}$.  This observation deserves a symmetry-based explanation.  Interestingly, for $\Delta=n\pi/4$ the lattice possesses additional symmetries that protect such FHOTI phases.  That is, without such underlying symmetries, the corner modes are unprotected and the system can be perturbed into the NI phase without closing the bulk gap, as indicated by the phase diagram [Fig.~\ref{square}(a)].

Specifically, for $\Delta=0$ (or, equivalently, any even multiple of $\pi/4$), the lattice obeys a particle-hole symmetry $CH(k,t)C=-H^{*}(-k,t)$ and inversion symmetry $\mathcal{I} H(k,t) \mathcal{I} = H(-k,t)$, where $C=\sigma_{z}$ and $\mathcal{I} = \sigma_{x}$. As such, in terms of the evolution operator, $CU_{F}(k_x,k_y)C=U_{F}^{*}(-k_x,-k_y)$ and $\mathcal{I}U_{F}(k_x,k_y)\mathcal{I}=U_{F}(-k_x,-k_y)$. For $\Delta=\pi/4$
and other odd multiples of $\pi/4$, the evolution operator satisfies $CU_{F}(k_x,k_y)C=U_{F}^{*}(\pi/a-k_x,-k_y)$ and $\mathcal{I}U_{F}(k_x,k_y)\mathcal{I}=U_{F}(\pi/a-k_x,-k_y)$.  With either set of symmetries, the lattice belongs to the $D$ class, which is associated with a $Z_{2}$ topological invariant \cite{Khalaf2018HOTI}.  These FHOTI phases are thus adequately described by the existing classification scheme for static HOTIs.

\begin{figure}
\includegraphics[width=1\linewidth]{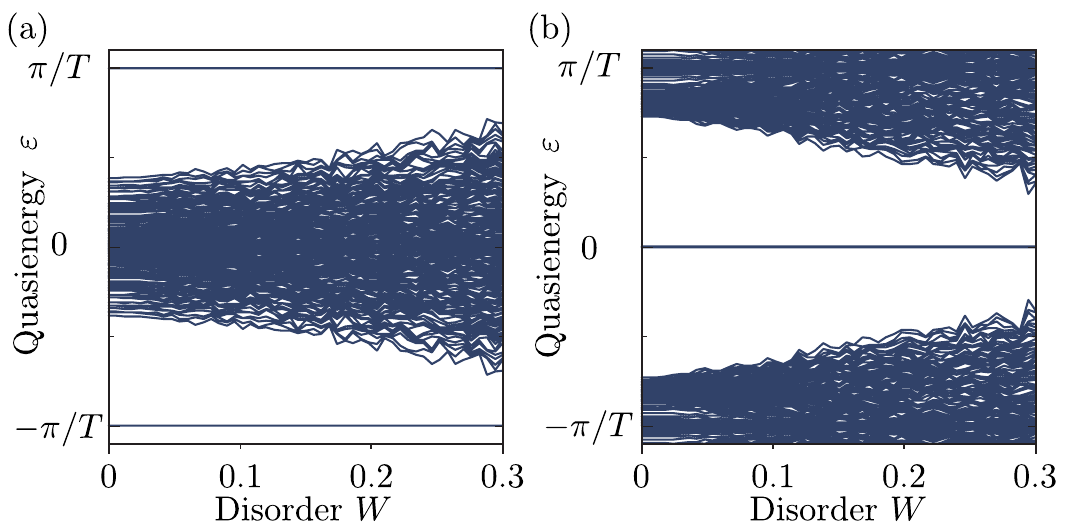}
\caption{Quasienergy spectra showing the robustness of the FHOTI corner modes against disorder.  The finite lattices are of the same shape as in Fig.~\ref{diagram} but with $8$ unit cells along each direction.  Each coupling is given by $\theta=\theta_{0}(1+D)$, where $D$ is uniformly distributed in $[-W,W]$.  The quasienergy spectra are plotted for varying disorder strength $W$ (with $D$ re-drawn for each $W$), with $\theta_{0}=0.9\pi$ and two different choices of detuning: (a) $\Delta=0$, and (b) $\Delta=\pi/4$.  With increasing disorder strength, the bulk gap narrows but the corner modes remain pinned at $\varepsilon = 0$ or $\pi/T$.}
\label{disorder}
\end{figure}

The particle-hole symmetry guarantees that the corner modes come in pairs with quasienergy $0$ or $\pi/T$, and the inversion symmetry further ensures that the paired corner modes are localized at different edges. So a single pair of corner modes cannot annihilate without bulk band closure.  To verify this, we study lattices with disorder in the coupling strength that preserves the particle-hole and inversion symmetries.  This is accomplished by setting $\theta=\theta_{0}(1+D)$, with $D$ randomly distributed in the range $-W<D<W$. As shown in Fig.~\ref{disorder}(a)--(b), both the $\pi$ corner mode (for $\Delta=0$) and the $0$ corner mode (for $\Delta=\pi/4$) are robust against the introduced disorder.  With increasing $W$, the bulk band broadens but the corner modes remain intact, with their quasienergies unaltered.  {In the Supplementary Material, we investigate the role of time-dependent disorder, showing that $2T$ perturbations can give rise to an interesting interplay between corner modes and chiral edge modes \cite{supp}.}

{\it Experimental proposal for FHOTIs in waveguide arrays.}---Optical waveguide arrays are an appealing platform for exploring Floquet topological phases.  Thus far, they have been used to realize Floquet Chern insulators and AFIs~\cite{Leykam2016waveguide, Maczewsky2017waveguide, Maczewsky2020waveguide, Mukherjee2017waveguide, Mukherjee2018waveguide, Mukherjee2020waveguide, Rechtsman2013waveguide}. Here, we present continuum simulations (not limited to the tight-band approximation) showing that the above FHOTI model can be realized in experimentally-realistic waveguide arrays.  By decreasing the distance between neighboring waveguides, one can achieve topological phase transitions from NI to AFI, and subsequently to the FHOTI phase.

The evolution of light in a waveguide system is governed by the Schr\"{o}dinger-like equation
\begin{equation}\label{eq4}
  i\partial_z\psi
  = -\frac{1}{2k_0}\nabla^{2}_{\bot}\psi-\frac{k_0\delta n(x,y,z)}{n_0}\psi,
\end{equation}
where $x,y$ are the transverse directions, $z$ is the axial direction, $\nabla_{\bot}^{2}=\partial_x^2+\partial_y^2$, $n_0$ is the background refractive index, $k_0=2\pi n_0/\lambda$, $\lambda$ is the operating wavelength in free space, and $\delta n(x,y,z)$ is the modulation in the refractive index \cite{Leykam2016waveguide, Mukherjee2020waveguide, Rechtsman2013waveguide}.  The axial coordinate $z$ acts as a synthetic time coordinate.  We take $n_0=1.473$ and $\lambda = 1550\,\textrm{nm}$.  The modulation function induces a bipartite square lattice with the unit cell configuration shown in Fig.\ref{diagram}(c), with both waveguides having clockwise helicity with helix radius $R = 4\,\mu\textrm{m}$, period $Z = 2\,\textrm{cm}$, and mean waveguide separation $d$.  Each waveguide has $\delta n=2.6\times10^{-3}$, with an elliptical cross section with major and minor axes of $9.8\,\mu\textrm{m}$ and $6.4\,\mu\textrm{m}$. All those parameters have been used in a previous experiment\cite{Noh2017waveguide}.

\begin{figure}
\includegraphics[width=1\linewidth]{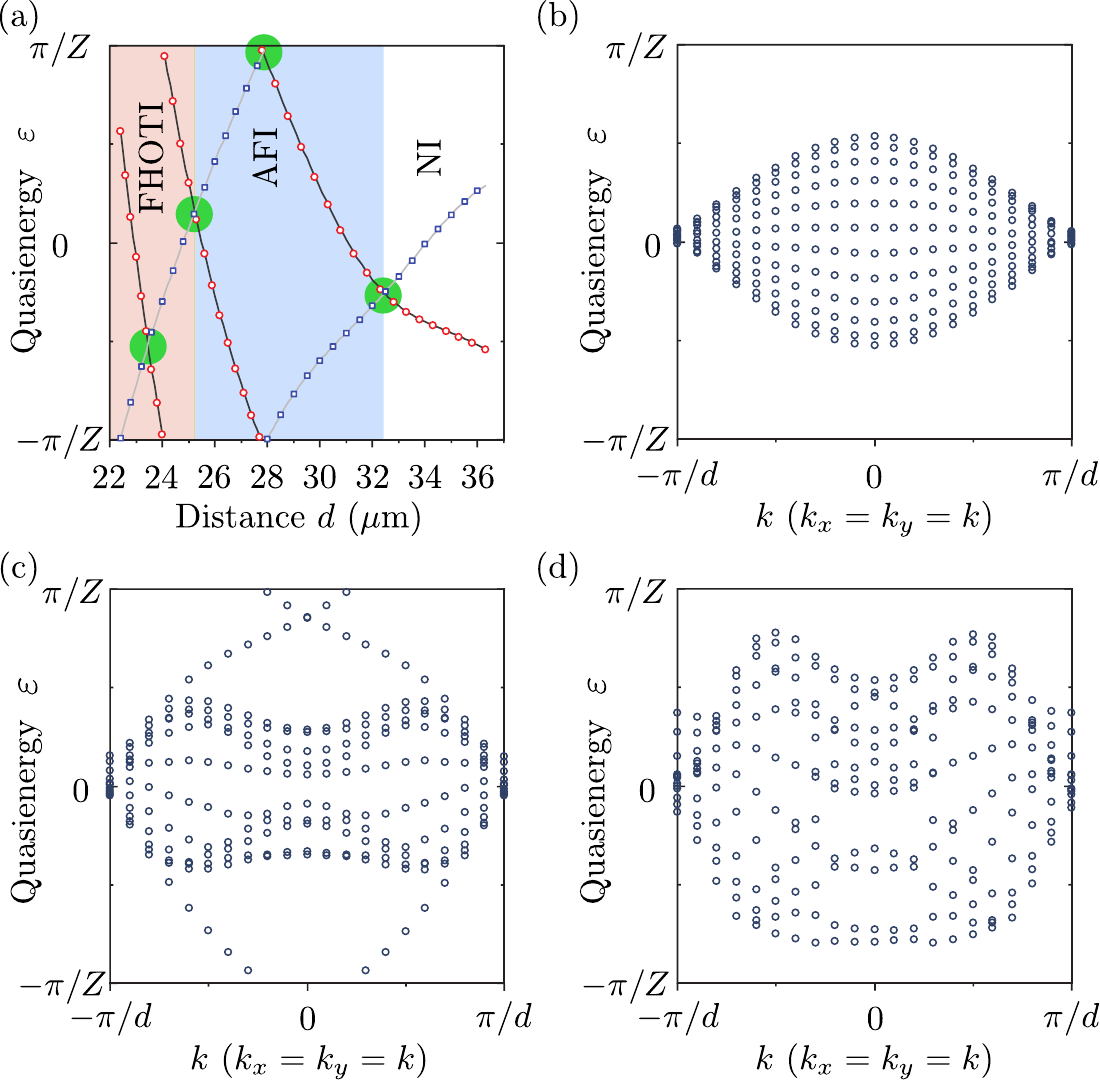}
\caption{Band diagrams for waveguide arrays with $\Delta=0$. (a) Band quasienergies at at $\Gamma$ (the center of the Brillouin zone) versus the inter-waveguide separation distance $d$.  The band-crossing points, marked by green circles, appear to correspond (in order of decreasing $d$) to the tight-binding model's critical coupling strengths $\theta \in \{\pi/4, \pi/2, 3\pi/4, \pi\}$ respectively [see Fig.~\ref{square}(a)].  (b)--(d) Band structures calculated for a semi-infinite strip that is periodic along the $(e_x, e_y)$ direction and $6$ unit cells wide along the $(e_x, -e_y)$ direction, for (b) $d=36\,\mu\textrm{m}$, (c) $d=28\,\mu\textrm{m}$ and (d) $d=23\,\mu\textrm{m}$. }

\label{bandstructure}
\end{figure}

The band structure of the waveguide array is numerically obtained by extracting the fundamental modes of waveguide, as described in Ref.~\onlinecite{Leykam2016waveguide}. We consider an array with zero detuning between the two sublattices (i.e., $\Delta = 0$).  For this case, the phase transition is predicted to occur at the $\Gamma$ point.  Fig.~\ref{bandstructure}(a) shows the calculated bulk band quasienergies at $\Gamma$, as a function of the waveguide separation $d$.  For $d \rightarrow \infty$, the lattice must be in the NI phase ($\theta \rightarrow 0$).  With decreasing $d$, we observe a band-crossing at $d \approx 32.4\,\mu\textrm{m}$, which should correspond to the NI-to-AFI transition at $\theta=\pi/4$ [Fig.~\ref{square}(a)].  The third band-crossing occurs at $d \approx 25.2\,\mu\textrm{m}$ and should correspond to the AFI-to-FHOTI transition at $\theta=3\pi/4$.

To verify the nature of these phases, Fig.~\ref{bandstructure}(b)--(d) shows the Floquet bandstructures for $d = 36\,\mu\textrm{m}$ (NI), $d = 28\,\mu\textrm{m}$ (AFI), and $d = 23\,\mu\textrm{m}$ (FHOTI), in a semi-infinite strip geometry.  As expected, topological edge modes are observed in the AFI case [Fig.~\ref{bandstructure}(c)].  The corner modes of the FHOTI phase cannot be seen here [Fig.~\ref{bandstructure}(d)], since the strip geometry lacks corners.

\begin{figure}
\includegraphics[width=1\linewidth]{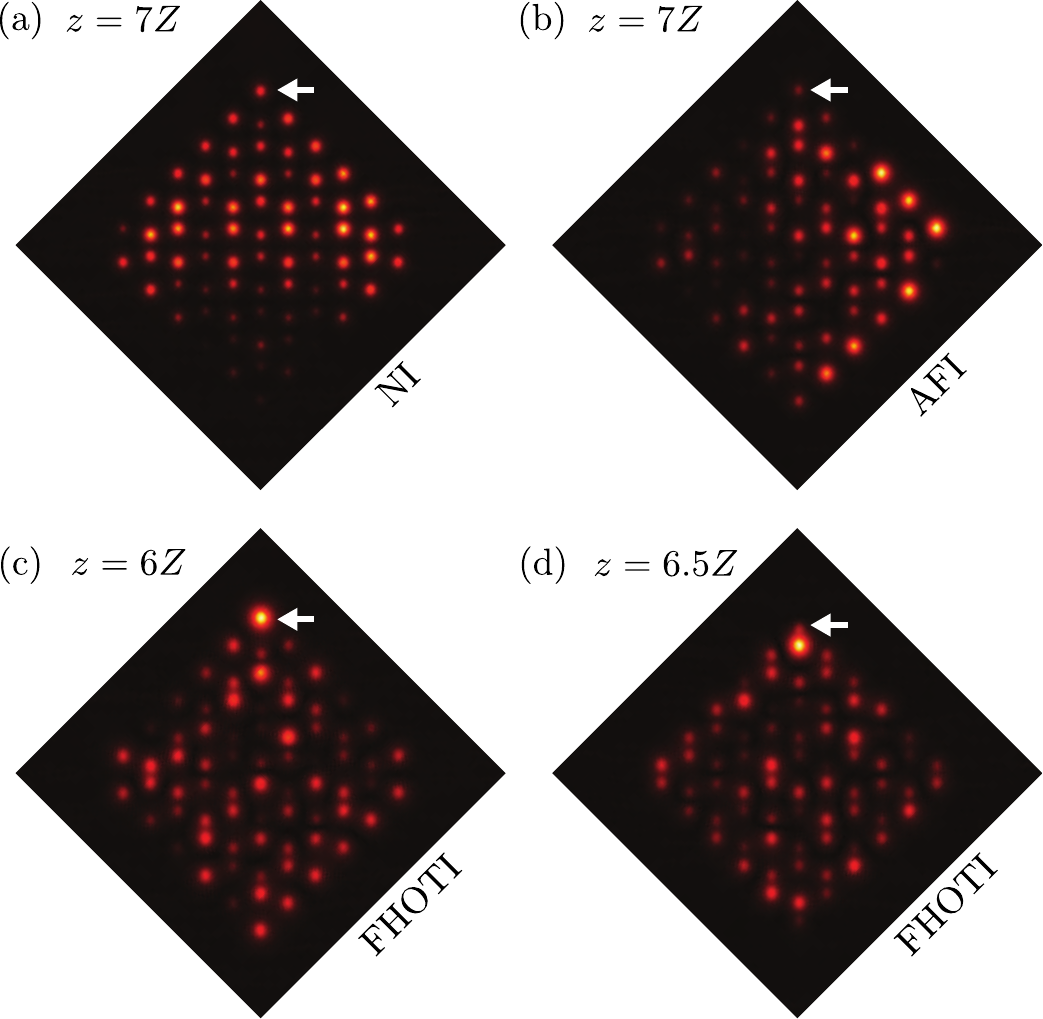}
\caption{Field intensity distributions in the optical waveguide array for different topological phases and propagation distances. (a) $d=36\,\mu\textrm{m}$ (NI phase) after $7Z$. (b) $d=28\,\mu\textrm{m}$ (AFI phase) after $7Z$. (c)--(d) $d=23\,\mu\textrm{m}$ (FHOTI phase) after (c) $6Z$ and (d) $6.5Z$.  The initially-excited corner waveguide is indicated by a white arrow. }
\label{simulationfield}
\end{figure}

Fig.~\ref{simulationfield} shows simulation results for a finite waveguide array, computed using the split-step Fourier method \cite{Leykam2016waveguide}.  At $z = 0$ (which corresponds to the maximal distance between the waveguides in the unit cell), a corner waveguide is excited.  The intensities are plotted at different final values of $z$.  For the NI, the initial corner excitation diffracts into the bulk [Fig.~\ref{simulationfield}(a)].  For the AFI, the corner excitation couples to chiral edge modes that propagate around a corner [Fig.~\ref{simulationfield}(b)].  For the FHOTI, the excitation remains localized, oscillating periodically between the few sites nearest to the corner.  In Fig.~\ref{simulationfield}(c)--(d), we plot the results at $6Z$ and $6.5Z$, for which the excitation is mostly confined to a single site (on the $A$ sublattice at $6Z$ and the $B$ sublattice at $6.5Z$). This interesting oscillation reflects the expected nontrivial microscopic dynamics of an edge at quasienergy $\pi/T$~\cite{Asboth2013pimode,Asboth2014pimode,Tong2013pimode,Cheng2019pimode}, but the observation here is even more interesting because it is due to a corner $\pi$ mode.  In the Supplementary Material \cite{supp}, we introduce a more complicated driving protocol that yields simultaneous $0$ and $\pi$ corner modes, which may be a feature of interest to the formation of time crystals \cite{Gong2018} and quantum information applications \cite{Gong2020}.

{\it Conclusion and discussion.}---In past studies, the engineering of a FHOTI phase has typically started from a Floquet topological insulator with gapless edge modes, and breaking a symmetry to open a gap in the edge modes.  The present model, by constrast, exhibits topological transitions between normal insulator, anomalous Floquet topological insulator, and FHOTI phases by varying system parameters while preserving lattice symmetry.  This bipartite lattice, with strictly non-negative nearest-neighbor couplings, appears to be the simplest setup known to date for realizing a FHOTI.  Variants of the model, based on honeycomb and triangular lattices rather than square lattices, can also realize FHOTI phases \cite{supp}.
We have shown using continuum simulations that the model can be realized in optical waveguide arrays of the sort that have previously been used to realize Floquet topological insulators experimentally.  This provides opportunities for studying the properties of these unusual topological phases, such as the interactions between corner modes and chiral edge modes.


\begin{acknowledgements}
J.G. acknowledges funding support by the Singapore Ministry
of Education Academic Research Fund Tier-3 (Grant No.
MOE2017-T3-1-001 and WBS No. R-144-000-425-592) and
by the Singapore NRF Grant No. NRF-NRFI2017-04 (WBS
No. R-144-000-378- 281).
We are grateful to L. Li, M. Umer, R. Bomantara, H. Xue and B. Zhang for helpful discussions.
\end{acknowledgements}

\clearpage
\onecolumngrid
\newcommand\rxout{\bgroup\markoverwith{\textcolor{red}{\rule[.5ex]{2pt}{.6pt}}}\ULon}
\newcommand\rxadd{\textcolor{red}}
\setcounter{equation}{0}
\setcounter{figure}{0}
\setcounter{table}{0}
\setcounter{page}{1}
\setcounter{section}{1}
\makeatletter
\renewcommand{\thesection}{\Roman{section}}
\renewcommand{\theequation}{S\arabic{equation}}
\renewcommand{\thefigure}{S\arabic{figure}}
\renewcommand{\bibnumfmt}[1]{[S#1]}
\renewcommand{\citenumfont}[1]{S#1}
\begin{center}
\textbf{\large Supplementary Materials}\end{center}

In these Supplementary Materials, we present a number of additional results.   In Sec.~\textcolor{blue}{\uppercase\expandafter{\romannumeral1}}, we show that a simple variation to the model in the main text, still retaining the two-band nature of the Hamiltonian, can yield  0 and $\pi$ corner modes simultaneously.   Sec.~\textcolor{blue}{\uppercase\expandafter{\romannumeral2}} presents simulation results showing how the coupling between the Floquet corner modes and chiral edge modes is affected by periodic perturbations of period $2T$.  In Sec.~\textcolor{blue}{\uppercase\expandafter{\romannumeral3}}, we discuss a way to understand the FHOTI phase by connecting it to a lower-dimensional topological insulator. Finally, Sec.~\textcolor{blue}{\uppercase\expandafter{\romannumeral4}} discusses FHOTI phenomena in periodically-driven honeycomb and triangular lattices, thus demonstrating the generality of our approach to realizing FHOTI phases.

\section{Coexistence of 0 and $\pi$ corner modes under a different driving protocol}
\label{s1}

Fig.~\ref{field0pi}(a) shows an example of periodic modulation to induce FHOTIs that support both $0$ corner modes and $\pi$ corner modes simultaneously, with $\Delta=0$.  Unlike the model in the main text, the driving protocol only has four steps.  We fix the coupling strength parameter $\theta$ at step 3 to $0.6\pi$.

The spectrum for a finite structure contains corner modes at both $\varepsilon = 0$ and $\varepsilon = \pi$, as shown in Fig.~\ref{field0pi}(b).  In Fig.~\ref{field0pi}(c)--(d), we plot the spatial profile of the $\pi$ corner mode, obtained by finding the eigenstates of the associated Floquet operator whose start time is set at different values.  (A different choice of the start time of the Floquet operator is equivalent to introducing some time evolution to the Floquet state within one period, hence the microscopic dynamics of Floquet states.)  We see that the corner mode consists of an oscillation between the A and B sublattices.  In Fig.~\ref{field0pi}(e)--(f), we show the spatial profile of the $0$ corner mode, which mainly occupies only the A sublattice as time evolves.

\begin{figure}
\includegraphics[width=0.6\linewidth]{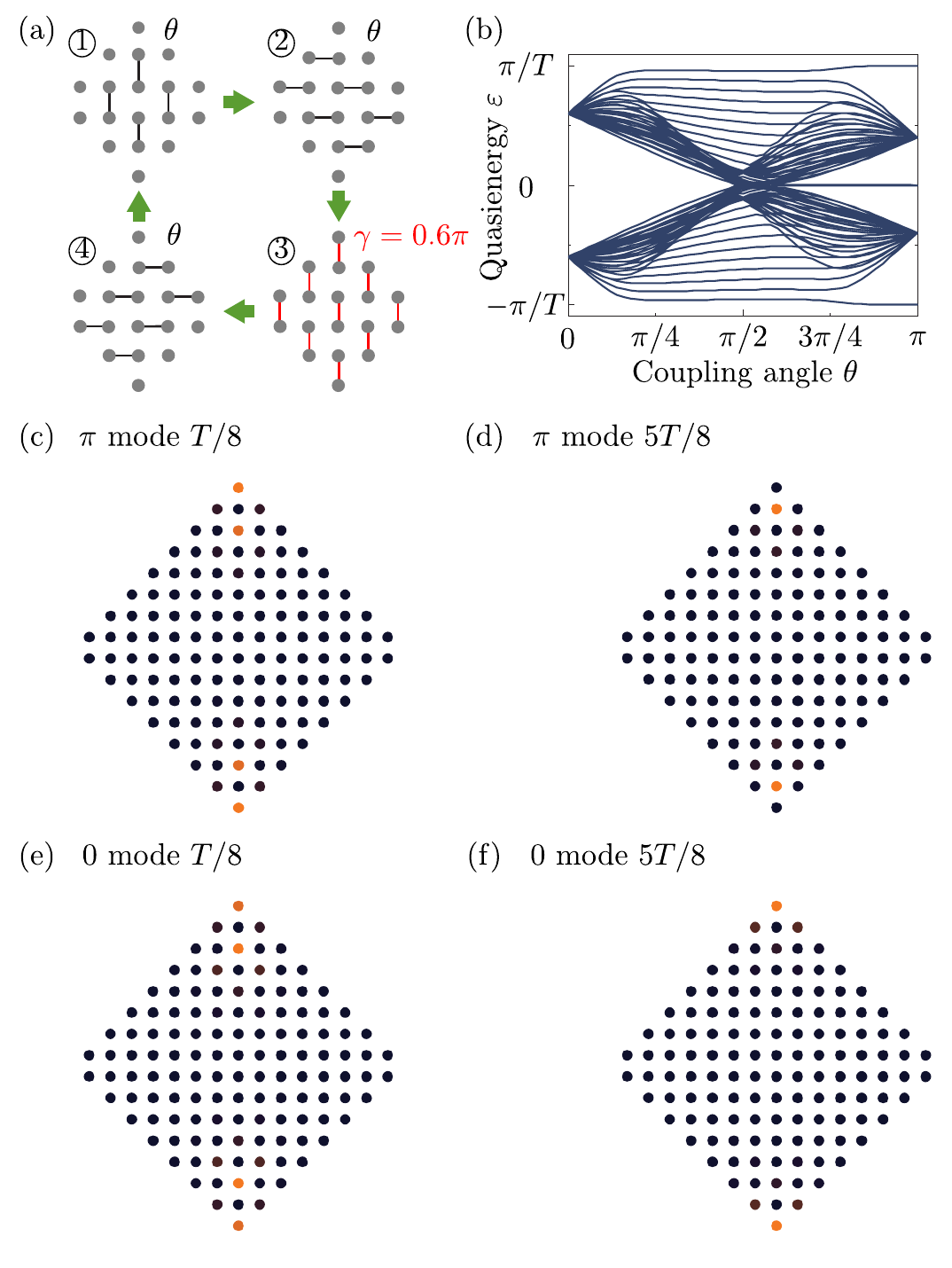}
\caption{A system that supports $0$ corner modes and $\pi$ corner modes simultaneously, with $\Delta=0$. (a) Driving protocol. The periodic modulation is composed of four steps and we fix the coupling strength as $0.6\pi$ in step $\textcircled{3}$. (b) Spectrum of a finite structure as a function of $\theta$ with $\Delta=0$. (c)--(d) The spatial profile of a $\pi$ corner mode for different initial time. (e)--(f) The spatial profile of a $0$ corner mode.  $\theta=0.9\pi$ for (c)--(f).}

\label{field0pi}
\end{figure}

\section{Coupling between 0 corner mode and chiral edge mode}
\label{s2}

From Fig.~2(d) in the main text, it is seen that the model hosts corner modes in the 0 gap and chiral edge modes in the $\pi$ gap for $\Delta=\pi/4$ and $\pi/2<\theta<\pi$.  Here we consider a scenario in which corner modes and chiral edge modes can be coupled by introducing perturbations of period $2T$.

There are many different ways to introduce $2T$-periodic perturbations.  We consider two cases:

In case 1, the $2T$ perturbation is introduced by changing the couplings $\theta$ in $T/16+2nT\leq t<T/16+(2n+1)T$ to $\theta-\delta$ and changing the couplings $\theta$ in $T/16+(2n+1)T\leq t<T/16+(2n+2)T$ to $\theta+\delta$. We choose $\theta=0.7\pi$ and study the spectrum of finite structure as we increase $\delta$. The results are shown in Fig.~\ref{corneredgecoupling}(a).  It is seen that with the increase of $\delta$, the corner modes are split. Fig.~\ref{corneredgecoupling}(b)--(d) show the spatial profile of a corner mode for $\delta=0,0.05\pi, 0.2\pi$, with the initial time to calculate the Floquet operator set at $9T/16$. From the shown results, it is observed that the corner modes become delocalized and finally are transformed into a propagating wave.

\begin{figure}
\includegraphics[width=0.6\linewidth]{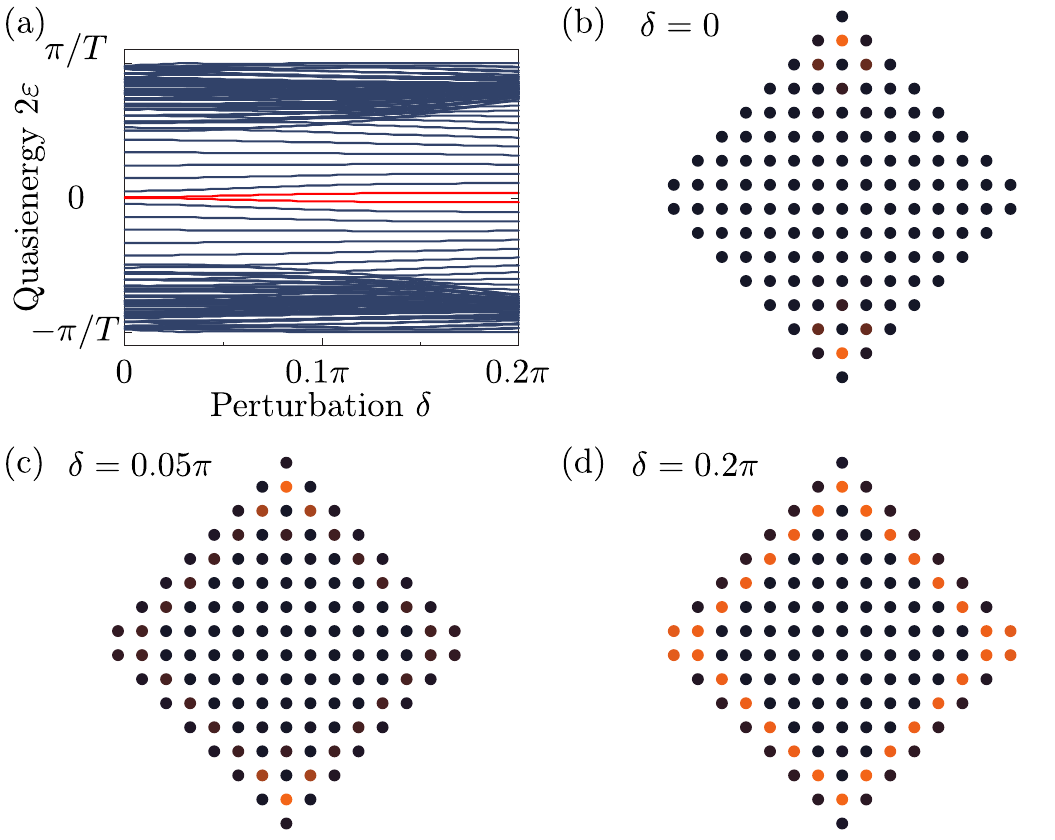}
\caption{Coupling between corner mode and chiral edge mode with $2T$ perturbation in case 1. (a) The spectrum for a finite structure as a function of $\delta$. (b)--(d) The spatial profile of the corner mode with $\delta=0,0.05\pi, 0.2\pi$. }

\label{corneredgecoupling}
\end{figure}

In case 2, the $2T$ perturbation is introduced by changing the couplings $\theta$ in $9T/16+2nT\leq t<9T/16+(2n+1)T$ to $\theta-\delta$ and changing the couplings $\theta$ in $9T/16+(2n+1)T\leq t<9T/16+(2n+2)T$ to $\theta+\delta$. We still choose $\theta=0.7\pi$ and then study the spectrum of a finite structure as we increase of $\delta$. The results are shown in Fig.~\ref{corneredgeuncoupling}(a). Different from case 1, here the corner modes survive but become corner modes embedded in the chiral edge continuum. These interesting observations are further confirmed by the shown Floquet state profiles in Fig.~\ref{corneredgeuncoupling}(b)--(d), where the corner modes are always localized (the initial time in our calculations is set at $T/16$).

\begin{figure}
\includegraphics[width=0.6\linewidth]{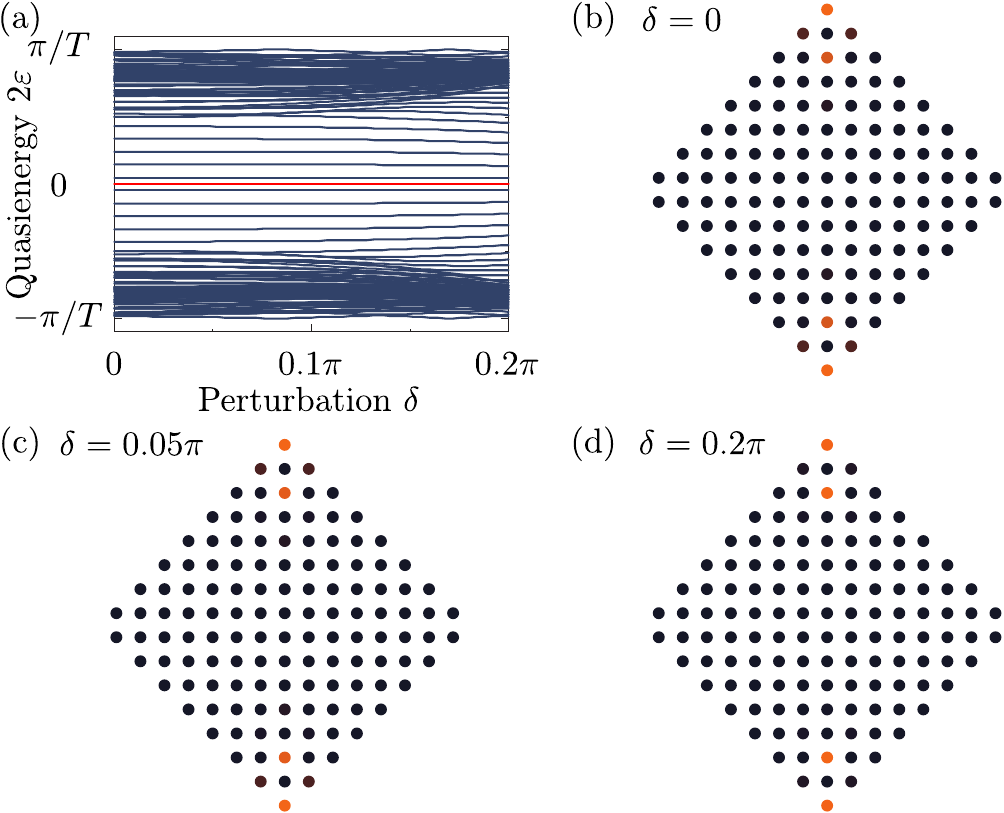}
\caption{Coupling between corner mode and chiral edge mode with $2T$ perturbation in case 2. (a) The spectrum for a finite structure as a function of $\delta$. (b)--(d) The spatial profile of the corner mode with $\delta=0,0.05\pi, 0.2\pi$.}

\label{corneredgeuncoupling}
\end{figure}

\section{FHOTIs in connection with a lower-dimensional topological insulator.}
\label{s3}

In this section, we offer more insights to the nature of the FHOTI phase. One way to understand HOTIs is to connect them to a lower-dimensional topological insulator \cite{Khalaf2018HOTI}. In our case, the FHOTIs can be connected with a 1D BDI insulator with chiral symmetry at specific quasimomentum $k_x$. To characterize this, let $U_F \equiv \mathrm{exp}\left[-iH_FT\right] = \cos\alpha+i \mathbf{n}\cdot\mathbf{\sigma} \sin \alpha$, where $\mathbf{n}=(n_x,n_y,n_z)$ is a unit vector, $\mathbf{\sigma}=(\sigma_x,\sigma_y,\sigma_z)$, and $H_F = \frac{\alpha}{T}\mathbf{n} \cdot \mathbf{\sigma}$ is the Floquet effective Hamiltonian.  Ensuring chiral symmetry of the periodically driven system amounts to ensuring that there exists an initial time $t_0$ such that the associated Floquet effective Hamiltonian has a chiral symmetry \cite{Asboth2013pimode,Asboth2014pimode,Tong2013pimode}. Such initial times for our system are found to be $T/16$ and $9T/16$. Here we specifically consider $t_0=T/16$, for which the $z$ component of the Floquet effective Hamiltonian is
\begin{equation}\label{eq3}
n_{z}\sin\alpha = \cos^{3}\theta\sin(4\Delta)+ \sin^{2}\theta\cos\theta\left[\sin(2k_x)+2\cos k_y\sin(k_x-2\Delta)\right].
\end{equation}
One can see that $n_z=0$ if $\Delta=n\pi/4$ and $k_x= m\pi+ 2\Delta$, where $m,n\in Z$. This indicates that, with special values of $\Delta$ and $k_x$, the effective Hamiltonian has the chiral symmetry $\sigma_{z}H_F\sigma_{z}=-H_F$.

For the above-mentioned special value of $k_x$, we can further treat our system as a 1D system along the $y$ direction. When $\Delta=n\pi/4$, the system has the chiral symmetry. The end mode of this 1D system is then protected by this chiral symmetry and the FHOTI phases can be connected with this 1D topological insulator.  This understanding is consistent with what we discussed in the main text. Indeed, in the main text,  we show only at $\Delta=n\pi/4$, the system has additional particle hole symmetry and inversion symmetry that protect the FHOTI phases.   So it is clear now that from both 1D perspective and 2D perspective, $\Delta=n\pi/4$ is seen to be a necessary condition to obtain FHOTIs.

Let us now explain the emergence of topological corner mode and provide topological invariants to characterize the FHOTIs. The topological corner modes are ensured by three conditions: (a) the system has chiral symmetry at special momentum values; (b) the winding number ($0$ gap or $\pi$ gap) is nonzero; and (c) the system is gapped along the edges forming the corner. Condition (a) and (b) promise that the system has edge mode pinned at quasi-energy $0$ or $\pi$ when we using periodic boundary condition along the $x$ direction and open boundary condition along the $y$ direction. Condition (c) promises that the corner modes cannot propagate along the edge.  To appreciate these conditions more we use $\Delta=0$ as an example.  Condition (a) is apparently met because the system has chiral symmetry at $k_{x}=0$.

For condition (b), we calculate a winding number as the topological invariant of the system.  This can be done by writing down the evolution operator in the special time-symmetric form\cite{Asboth2014pimode}
\begin{equation}\label{eqs1}
  U(T,k_{x}=0)=\sigma_{z}F^{\dagger}\sigma_{z}F,
\end{equation}
where
\begin{eqnarray}
  F &=& e^{i\frac{\theta}{2}((\cos 2k_y)\sigma_x+(\sin 2k_y)\sigma_y)}\cdot e^{i\theta((\cos k_y)\sigma_x+(\sin k_y)\sigma_y)}\cdot e^{i\frac{\theta}{2}\sigma_x}  \nonumber \\
  &=& \left(\begin{array}{cc}
           a(k_y) & b(k_y) \\
            c(k_y) & d(k_y)
          \end{array}
  \right).
\end{eqnarray}
For the model detailed in the main text, there is only one gap at $\pi$ as shown in Fig.~2(b). The existence of edge modes in this gap can be  determined by the winding number based on the submatrix $d(k_y)$ defined above.  That is, the topological invariant to characterize the $\pi$ corner mode is given by the winding number defined by the complex function $d(k_y)$ when we vary $k_y$ from $-\pi$ to $\pi$\cite{Asboth2014pimode}:
\begin{equation}\label{eqs3}
  \upsilon_\pi=\upsilon(d).
\end{equation}

In Fig.~\ref{topologicalinvariant}(f), we present results of this winding number $v_\pi$ associated with the $\pi$ gap,  for different choices of $\theta$.  It is seen that $v_\pi=0$ for $\theta<\pi/4$, $v_\pi=1$ for $\pi/4<\theta<3\pi/4$ and $v_\pi=2$ for $\theta>3\pi/4$. These nonzero winding numbers determine the number of $\pi$ modes the system can support at $k_x=0$, if we calculate the band structure of a strip geometry, periodic along the $x$ direction and finite along the $y$ direction as shown in Fig.~\ref{topologicalinvariant}(a).  Fig.~\ref{topologicalinvariant}(d) show the band structure with $\theta=0.8\pi$, with $\pi$ modes at $k_x=0$.

Condition (c) is that the system is gapped along the edges forming the corner. Those edges are along $(e_x,e_y)$ and $(e_x,-e_y)$ direction as shown in Fig.~\ref{topologicalinvariant}(c). It is found that the system has gapless edge mode for $\pi/4<\theta<3\pi/4$ and is gapped for $\theta>3\pi/4$. Examples are shown in Fig.~\ref{topologicalinvariant}(e) and Fig.~\ref{topologicalinvariant}(g) for $\theta=0.8\pi$ and $\theta=0.6\pi$. The final conclusion is that the system supports topological corner modes with $\theta>3\pi/4$.   These results are all consistent with our observations made in the main text.

\begin{figure}
\includegraphics[width=0.6\linewidth]{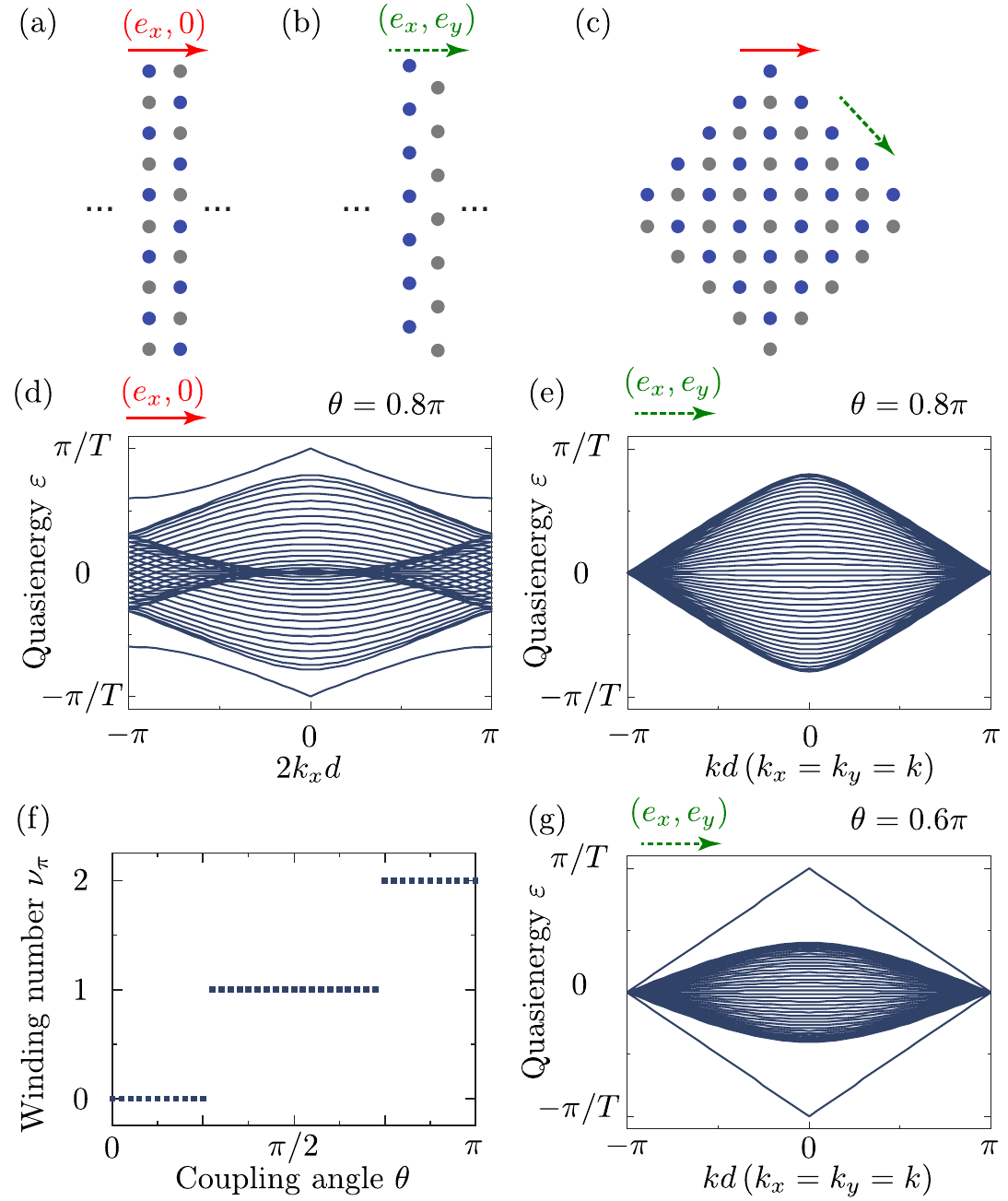}
\caption{Schematic for a strip geometry and finite structure. (a) A strip geometry which is periodic along the $x$ direction and finite along the $y$ direction. (b) A strip geometry which is periodic along $(x,y)$ direction and finite along $(x,-y)$ direction. (c) A finite structure. (d) Band structure corresponding to strip geometry in (a) with $\theta=0.8\pi$. (e) Band structure corresponding to strip geometry in (b) with $\theta=0.8\pi$. (f) Winding number in $\pi$ gap as a function of $\theta$ with $\Delta=0$. (g) Band structure corresponding to strip geometry in (b) with $\theta=0.6\pi$.}

\label{topologicalinvariant}
\end{figure}

\section{FHOTIs in a driven honeycomb lattice and a driven triangle lattice}
\label{s4}

To further show that our proposal works more generally than what is discussed in the main text, here we consider extensions to other driven lattices, using the same notation as in the main text.  Fig.~\ref{honeycomb} shows how one may generate FHOTIs in  a driven honeycomb lattice. Fig.~\ref{honeycomb}(a) illustrates  coupled ring resonators with coupling strength $\theta=\pi$.  In this case, all the rings are "uncoupled" and the bulk ring has different resonance frequency with the edge ring and corner ring. When $\theta$ is slightly away from $\pi$, the edge rings support edge modes that are localized at the edge. They form a Su-Schrodinger-Heeger model and support corner modes at the up corner and down corner. Fig.~\ref{honeycomb}(b) is the corresponding tight binding model. The phase diagram is presented in
Fig.~\ref{honeycomb}(c).  It is indeed seen that this mode also supports FHOTIs. The spectrum as a function of $\theta$ for a finite structure with $\Delta=0$ is shown in Fig.~\ref{honeycomb}(d).  One sees that for $2\pi/3<\theta<\pi$, the system supports corner modes in the 0 gap. Fig.~\ref{honeycomb}(e)--(f) confirm that those FHOTIs are robust against disorders introduced to the coupling strength.

\begin{figure}
\includegraphics[width=0.6\linewidth]{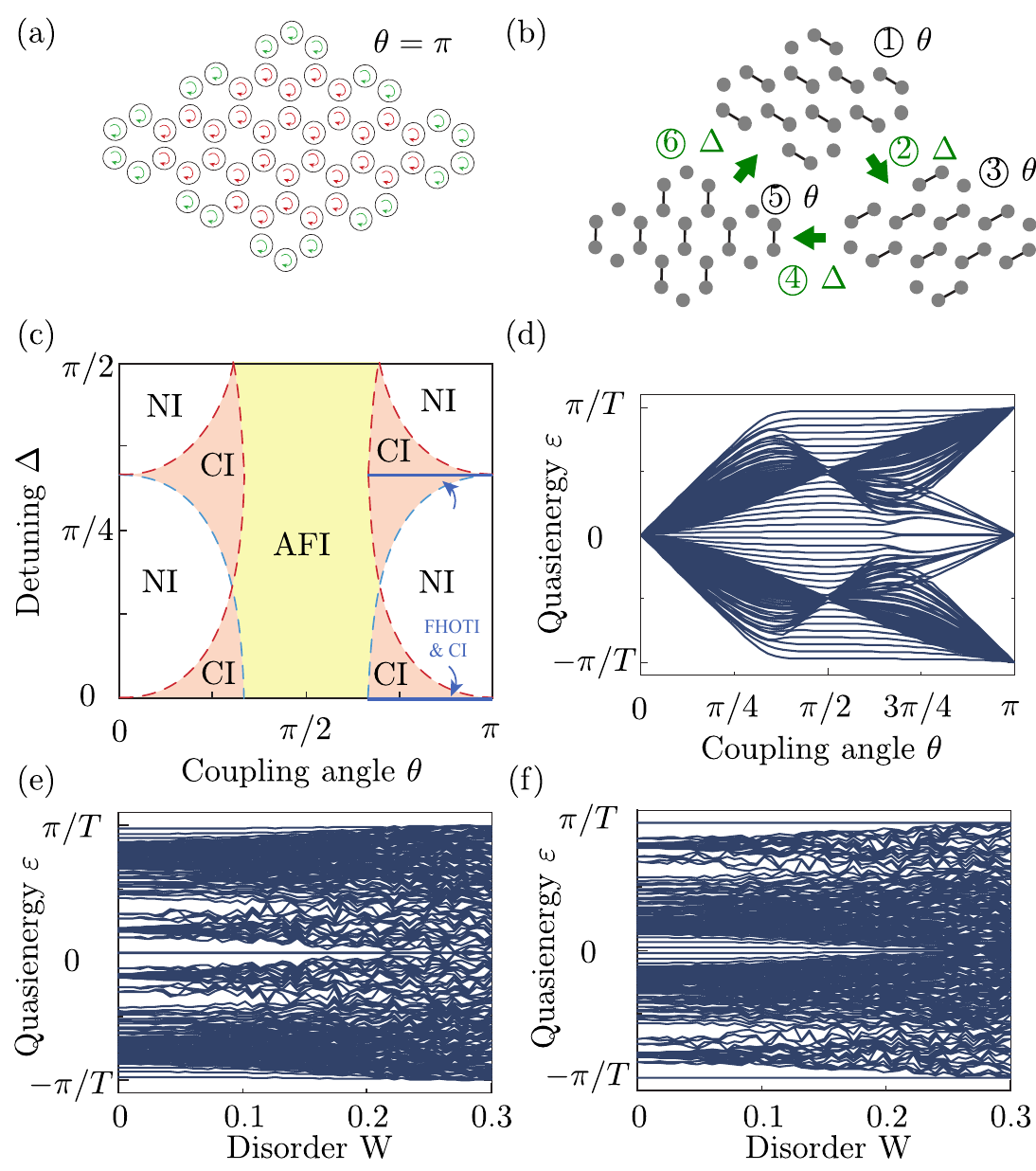}
\caption{A tight binding model and the associated phase diagram for couple ring resonators (CRRs) arranged in a honeycomb lattice structure . (a) The honeycomb CRRs lattices with a finite structure. The wave evolution with coupling strength $\theta=\pi$ is illustrated. The bulk modes (red) and edge modes (green) are all propagated locally but have different frequency. When tuning $\theta$ away from $\pi$, the edge modes are coupled and form a SSH model. Topological zero corner mode is supported at the obtuse corner. (b) Driving protocol. The periodic time modulation is composed of 6 steps including changes to three nearest neighbor coupling terms $\textcircled{1}\textcircled{3}\textcircled{5}$ and three on site difference terms $\textcircled{2}\textcircled{4}\textcircled{6}$. (c) Phase diagram as a function of $\theta$ and $\Delta$. Blue (red) dashed lines are the phase transition points at $\Gamma$ (K or K') point. (c)--(d) Spectrum as a function of coupling strength $\theta$ for a finite structure with 5 unit cells along each direction. $\Delta=0$. (e)--(f) Robustness against disorder. $\theta_{0}=0.8\pi$ for (e)--(f). $\Delta=0$ for (e), $\Delta=\frac{\pi}{3}$ for (f).}

\label{honeycomb}
\end{figure}

Next we consider the generation of FHOTIs in a driven triangle lattice and the results are shown in Fig.~\ref{triangle}. Specifically, Fig.~\ref{triangle}(c) is CRRs in a triangle lattice with $\theta=\pi$ and Fig.~\ref{triangle}(a) is the corresponding tight binding model. Different from the square lattice and honeycomb lattice situations, here each ring is replaced by two sites. The phase diagram for this model is plotted in Fig.~\ref{triangle}(b).  It is seen again that this model supports FHOTIs and can even simultaneously support 0 corner modes and $\pi$ corner modes. Those results are confirmed by the spectrum for a finite structure in Fig.~\ref{triangle}(d)--(f). The robustness of those corner modes against disorder introduced to the coupling strength is further confirmed by results shown in Fig.~\ref{triangle}(g)--(h).

\begin{figure}
\includegraphics[width=0.6\linewidth]{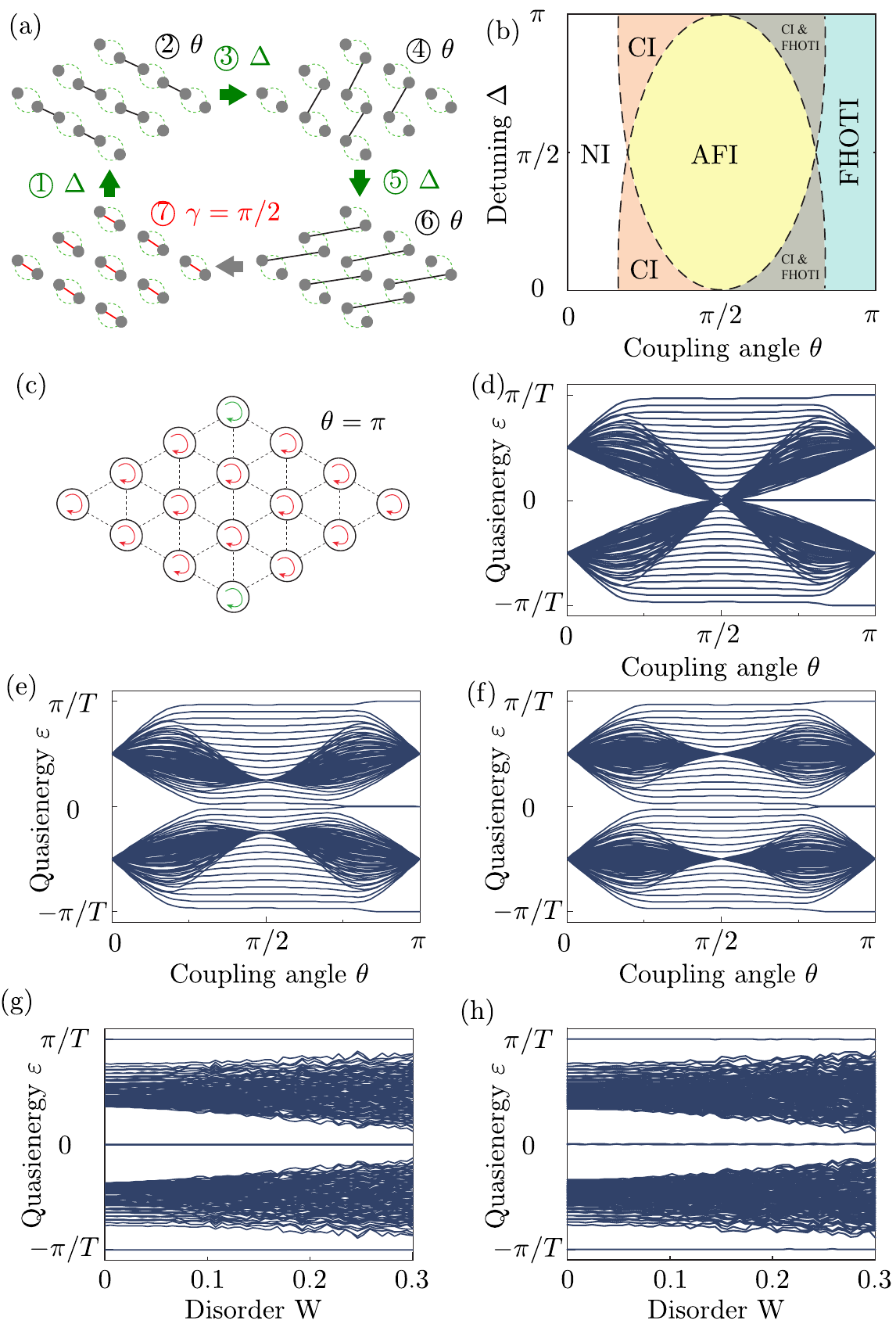}
\caption{A  tight-binding model and the associated phase diagram for CRRs arranged in a triangular lattice. (a) Driving protocol. The periodic time modulation is composed of 7 steps including changes to four nearest neighbor coupling terms $\textcircled{2}\textcircled{4}\textcircled{6}\textcircled{7}$ and three on-site difference terms \textcircled{5}. Different from our previous models, the couplings strength in step $\textcircled{7}$ is fixed at $\frac{\pi}{2}$. (b) Phase diagram as a function of $\theta$ and $\Delta$ (there dashed lines are the phase transition points), featuring again the emergence of FHOTIs.   (d)--(f) Spectrum as a function of coupling strength $\theta$ for a finite structure with 8 unit cells along each direction. $\Delta=0$ for (d), $\Delta=\frac{\pi}{4}$ for (e) and $\Delta=\frac{\pi}{2}$ for (f). (g)--(h) Robustness against disorder introduced to the coupling strength. $\theta_{0}=0.9\pi$ for (g)--(h). $\Delta=0$ for (g), $\Delta=0.3$ for (h).}

\label{triangle}
\end{figure}

\end{document}